\documentstyle[12pt]{article}
\newlength{\myleftmargin}
\newlength{\paperwidth}
\begin{document}
\thispagestyle{empty}
\baselineskip 18pt
\begin{flushright}
   {\tt DPNU-94-49 \\
        November 1994}
\end{flushright}
  \vspace{0.1em}
\begin{center}
   {\Large {\bf The $F/D$ Ratios of Spin-Flip Baryon Vertex}}
\vskip 1mm
 {\Large {\bf in $1/N_c$ Expansion}}
\footnote[1]{The main part
  of this paper was presented at the International Bogolubov
Symposium,
   held on 21--24, August 1994, Dubna and at the Annual Meeting of
Japanese
    Physical Society held on 28, September--1, October, Yamagata.\\
\hskip 3mm e-mail adresses: $<$takamura@eken.phys.nagoya-u.ac.jp$>$,
   $<$sawadas@eken.phys.nagoya-u.ac.jp$>$,
 and $<$kitakado@eken.phys.nagoya-u.ac.jp$>$}

\end{center}
  \vspace{0.3em}
\begin{center}
  {\sc Akira~~Takamura},~~~{\sc Shoji~~Sawada},~~~{\sc
Shinsaku~~Kitakado}
\end{center}
\begin{center}
  \sl{Department of Physics, Nagoya University,}\\
  \it{Nagoya 464-01, Japan}
\end{center}
\vskip 1.5em  
{\large {\bf Abstract~:}}
 We calculate the $F/D$ ratios of spin 1/2 baryon vertex for both
 the
 non-relativistic quark model and the chiral soliton model with
 arbitrary
  number of color degrees of freedom $N_c$ and examine the results
 in terms of
   the consistency
   condition  approach for the baryon vertices recently developed by
 Dashen,
 Jenkins and Manohar from the viewpoint  of QCD.
\par
 We show that the $1/N_c$ corrections have two different origins, i.e.
 one is
 from the baryon states or baryon wave functions and the other from
 the
 vertex operators.
 Although in the limit of $N_c \to \infty$ the $F/D$ tends to 1/3
 in  all
  models, the $1/N_c$ expansion of $F/D$ ratio
 does not  converge for
  $N_c=3$ in the chiral soliton model in contrast
 to the non-relativistic
  quark model.
  \vskip 0.5em  
\section{Introduction}
\hspace*{\parindent}
The non-relativistic quark model(NRQM)  has
 successfully described the hadron
phenomena in various
 aspects and played a crucial role in the way
 of establishing QCD.
 However, the NRQM description for the hadron
 states has not been derived
  convincingly from QCD.
This is mainly due to the nonperturbative nature
 of QCD in the low energy
regions. In order to overcome this difficulty,
 in 1974,  't Hooft introduced
a hidden small expansion parameter of  QCD,
 $1/N_c$$\cite{'t Hooft}$.
 In 1979 Witten$\cite{Witten}$ applied the
$1/N_c$ expansion method to
  baryon and suggested that  in the large
 $N_c$ limit
the baryons are realized as  solitons of
meson fields because
the low energy effective Lagrangian does
 not contain baryons due to their large
mass of order $N_c$.
 From the early 1980's the Skyrme's conjecture
 that baryons are the soliton
  of the nonlinear chiral Lagrangian for the
  chiral fields has been
   revived\cite{Nappi}. By quantizing the
collective modes
  of rotations of soliton in the spin-isospin
 space we obtain the ``spinning"
   soliton  states which  corresponds
  to baryons and various properties of baryons
 are well reproduced in terms of
    the parameters of meson sectors.
  Most important characteristics of the chiral
 soliton will be the hedgehog
 structure which gives  ${\bf I} +{\bf J}=0$.

On the basis of the $1/N_c$ expansion method
we have been  able to understand
 many of the $qualitative$ properties of mesons
 and baryons,
 the $quantitative$ properties of hadrons, however,
 could not be derived
  up to very recently.
 The first quantitative relation for hadron
 phenomena was obtained from the
 consistency condition for the baryon vertex
 in the pion-nucleon scattering
  amplitude which was derived by  Gervais and
 Sakita$\cite{Gervais}$ on the
  basis of the large $N_c$ expansion and the
strong coupling theory.
Recently Dashen, Jenkins and Manohar$\cite{Dashen}$
 have developed this
consistency condition approach from the view of QCD
 and derived an induced
algebra with respect to the spin-flavor symmetry.
 On the basis of the
consistency conditions they have analysed the
 large $N_c$ behaviors of
 baryon vertices in terms of model-independent
 method and
examined  the vertices in  the NRQM  and the
 static $SU(2)$
 chiral soliton model(CSM).
Furthermore they have obtained 1/3 for the
 large $N_c$ limit of $F/D$
 ratio  from the consistensy conditions
 model-independently.

In this paper we  calculate the $F/D$
 ratios for arbitrary value of color
degrees of freedom $N_c$ for both the
 NRQMs
with flavor $f=2$ and $f=3$ using $ SU(4)$
 and $ SU(6)$ symmetries respectively  and
 for the $SU(2)$ and $SU(3)$ CSMs.

 We pay our attention to the fact that the
 $1/N_c$ corrections have two
 different  origins, i.e. one
 from the baryon states or baryon wave
 functions and the other from the
  vertex operators.

We compare the obtained results of $F/D$
 ratios
by using magnetic moments of proton and
neutron  and confirm that in all models
 the limiting value  of $F/D$ ratio of
 spin-flip vertex is 1/3.
We show that in the case of $SU(3)$,
 the magnetic moment of proton and neutron
 depend on how the real baryon states with
 $N_c =3$  are  extrapolated to
large  $N_c$ baryon states,
 the large $N_c$ limit,
but the $F/D$ ratio depends only on group
 structure of the baryon multiplet
  and not on the way of extrapolation of
 baryon states.
 We found also that although in the limit
 of $N_c \to \infty$ the $F/D$ ratio
  tends to 1/3 in  all
  models, the $1/N_c$ expansion of $F/D$
 ratio does not  converge for
  $N_c=3$ in the $SU(3)$ CSM in contrast
 to the NRQMs.

\section{The $F/D$ Ratio from $SU(4)$ and
 $SU(6)$ Symmetric Quark Model}
\hspace*{\parindent}
In the NRQMs with the $SU(4)$ and $SU(6)$
 symmetries
 the spin 1/2 baryons are given
by the completely symmetric representation
 with respect to spin and flavor
which are represented by the Young diagrams
 with  the first row of
length $N_c =2k+1$ ($k=0,1,2,...)$ only and
 have dimensions
 respectively the states in the
 $\bf {}_6 H_{N_c}$   and $\bf {}_4 H_{N_c}$,
 where  ${}_n H_r$ is a repeated
 combination ${}_n H_r = {}_{n+r-1} C_r =
 (n+r-1)!/r!(n-1)!$.

 \par
In the $SU(4)$ symmetric model the magnetic
 moment of baryon $B$ is given by
\[(\mu_B)_{SU(4)} =
 \left(
 \begin{array}{ccc}
  {\bf 15}&{\bf  {}_4 H_{N_c}}&{\bf {}_4 H_{N_c}}^* \\
  \sigma_3 Q& B& {\bar B}
 \end{array}
 \right) \mu \]
 where $\sigma_3$ represents spin up or
 down and $Q =( \tau_3 + 1)/2$ the
 charge of the nucleon and $\mu$ is an
unknown constant to the
 leading order in $1/N_c$.
  It is noted here that
isoscalsr part of the magnetic moment dose
 magnetic moment. But since
in the $SU(4)$ symmetry,
 constructed by the complete
both of the  isoscalar part and isovector
part of the spin-flip vertex are
generators which belong to the same $ SU(4)$
 supermultiplet {\bf15}. \par
Similarly in the $SU(6)$ symmetric model the
 magnetic moment of spin
1/2 baryon $B$ is given by
\[(\mu_B)_{SU(6)} =
 \left(
 \begin{array}{ccc}
  {\bf 35}& {\bf {}_6 H_{N_c}}&{\bf {}_6 H_{N_c}}^* \\
  \sigma_3 Q& B& {\bar B}
 \end{array}
 \right) \mu \]
 Here $Q=\lambda_3 /2+ \lambda_8/2\sqrt{3}$
is the charge operator and
  the isoscalar part and  isovector
part of the spin-flip vertex
  belong to the same supermultiplet
{\bf 35} of $SU(6)$.
\par
Both of the $SU(4)$ and $SU(6)$
symmetric  models  give the same magnetic
 moment for proton and neutron for arbitrary $N_c$:
\begin{eqnarray}
(\mu_p)_{SU(4),SU(6)} & = & (k+2)\mu \label{fdp} \\
(\mu_n)_{SU(4),SU(6)} & = & -(k+1)\mu.\label{fdn}
\end{eqnarray}
These results are already derived
by Karl and Paton$\cite{Karl}$.
Most general flavor octet vertex
constructed from spin 1/2 baryon $B$ is
given by$\cite{Dashen}$
\begin{eqnarray}
 {\cal M}{\rm Tr} [ {\bar B}T^a B] +
{\cal N}{\rm Tr} [{\bar B}B T^a],
 \end{eqnarray}
where $T^a$ a flavor octet matrix.
The $F/D$ ratio of flavor octet vertex is given by
\[ \frac{F}{D} =
\frac{{\cal M}-{\cal N}}{{\cal M}+{\cal N}}. \]
 By the use of this $F/D$ ratio the magnetic
moments of nucleons are given by
 \begin{eqnarray}
\mu_p = \mu_F + \frac{1}{3} \mu_D  \label{fdp1} \\
\mu_n = -\frac{2}{3} \mu_D ,\label{fdn1}
\end{eqnarray}
where $\mu_D$ and $\mu_F$ are the $D$ and $F$
 type contributions to the
magnetic moment of baryons.
For arbitrary $N_c$ we calculate
the ${\cal N}/{\cal M}$ ratio and $F/D$ ratio
for the ``baryon" with spin 1/2 and the obtained results are
\begin{eqnarray}
& & \left(\frac{\cal N}{\cal M}\right)_{SU(4),SU(6)}
 = \frac{N_c-1}{2(N_c+2)} \\
& & \left(\frac{F}{D}\right)_{SU(4),SU(6)}
=\frac{N_c + 5}{3(N_c + 1)}.
 \label{fdsu}
\end{eqnarray}
\hspace*{\parindent}
The same result can be derived from
the magnetic moments of nucleons
given by (\ref{fdp}), (\ref{fdn}), (\ref{fdp1}) and (\ref{fdn1}).
It is noted that in the case of NRQMs with
$SU(4)$ and $SU(6)$ symmetries the
coefficient $\mu$ of magnetic moment
 is different depending  on how the
large $N_c$ baryon states are defined.
However the $F/D$ ratio does not
depend on the way of extrapolation to
large $N_c$ baryons.

  In the $SU(4)$ and $SU(6)$ NRQMs
taking the limit of $N_c \to \infty$
   the $F/D$ ratio tends to 1/3. For
  $N_c =1$ the $F/D$ ratio  becomes
1 reflecting the fact that the baryons are
   quarks themselves for $N_c =1$.

\section{The $F/D$ Ratio in $SU(2)$ and
$SU(3)$ Chiral Soliton Model}
\hspace*{\parindent}
In the $SU(2)$ CSM the spin 1/2 baryon
state is represented
 by the elements of $SU(2)$ matrix
in the fundamental representation of $SU(2)$
 which is independent of color degrees of
freedom $N_c$$\cite{Adkins}$.
In this model the isoscalar part
and the isovector part of magnetic moment
 of baryon have
distinct origins. That is, the isovector part is space
integral of conserved isovector
current which is the Noether current reflecting
the symmetry of the chiral Lagrangian
and is of order $O(N_c)$. On the other hand
 the isoscalar part
 comes from the baryon number current which is a
topologically conserved current and of
order $O(1/N_c)$ in the $1/N_c$
 expansion.
Therefore the magnetic moment of the spin 1/2 baryon
is given by
\[
 (\mu_B)_{SU(2) \; CSM} =
  \left(
  \begin{array}{ccc}
  {\bf 3}& {\bf 2} & {\bf 2} \\
   \tau_3 & B& {\bar B}
  \end{array}
  \right)
  \left(
  \begin{array}{ccc}
   {\bf 3}& {\bf 2}& {\bf 2} \\
   \sigma_3& B& {\bar B}
  \end{array}
  \right) \mu^{I=1}  +\cdots \label{mub}
\]
where $\tau^3 /2 = Q^{I=1}$ is the isovector
part of charge. The  ellipsis in (\ref{mub})
 denotes contributions from time derivative
of dymamical variables of the
 spin-isospin rotation of chiral soliton in
which the isoscalar part of
 magnetic moment given by the topological or
baryon number current is contained.  \begin{eqnarray}
(\mu_p )_{SU(2) \; CSM} & = \frac{1}{2}\mu_{I=1} + \cdots,  \\
(\mu_n )_{SU(2) \; CSM} & = -\frac{1}{2}\mu_{I=1} + \cdots.
\end{eqnarray}
Then the $F/D$ ratio in the $SU(2)$ CSM is
\begin{equation}
\left(\frac{F}{D}\right)_{SU(2) \; CSM}= \frac{1}{3} +\cdots
\end{equation}
\hspace*{\parindent}
In the $SU(2)$ CSM the
isovector part is given by the vector
current of soliton which is a Noether
current, but the isoscalar part
denoted by the ellipsis is
 given by the baryon number
current which is of topological origin.
That is to say, there is no direct
relation between the isovector and the
 isoscalar part. In the static limit
there is no isoscalar part of the
 magnetic moment and the $F/D$ ratio
takes the large $N_c$ limiting value 1/3
 irrespective of $N_c$.
  \par
On the other hand, in the $SU(3)$ CSM the spin 1/2 baryon
states are
represented by the $SU(3)$ matrix
of the ${\bf (1,k)}={\bf (1+k)(3+k)}$
  dimensional representation  where $k=(N_c -1)/2$.
  ${\bf (1,k)}$ denotes the representation
of $SU(3)$ with the Young diagram
   which  has the first row of length $k+1$
and the second row of length $k$.
In the case $N_c=3$ this is the octet or the
regular representation of $SU(3)$.

In  contrast to the $SU(2)$ CSM, in the $SU(3)$ chiral
soliton model the magnetic moment
of spin 1/2 baryon $B$ is given
 by$\cite{Nappi}$
\[
 (\mu_B)_{SU(3) \; CSM} = \sum_n
  \left(
  \begin{array}{ccc}
  {\bf 8}& {\bf (1,k)}& {\bf (1,k)}_n^* \\
   Q& B& {\bar B}
  \end{array}
  \right)
  \left(
  \begin{array}{ccc}
  {\bf  8}&{\bf  (1,k)}& {\bf (1,k)}_n^* \\
   \sigma_3& B& {\bar B}
  \end{array}
  \right) \mu +\cdots
\]
where the summation over $n$ means
two orthogonal states of baryons of
$\bf (1,k)$
 representation and the ellipsis denote
corrections from the time derivative
  of the $SU(3)$
matrix valued dynamical variable $A(t)$
describing the ``rotations" in
spin-flavor space and contains the higher
order term of $1/N_c$ expansion.
The results for the magnetic moments are
\begin{eqnarray}
(\mu_p)_{SU(3) \; CSM} & = & \frac{k+3}{3(k+4)}
 \mu  + \cdots \\
(\mu_n)_{SU(3) \; CSM} & = & \frac{k^2+5k+3}{3(k+2)(k+4)}
 \mu + \cdots.
\end{eqnarray}
There is a difference in the magnetic moments
of nucleons in the chiral soliton
 models between flavor 2 and 3 contrary to the NRQMs.
 This comes from the fact
 that the nucleon states belong to
the fundamental representation ${\bf 2}$
 in the $SU(2)$ soliton irrespective
of $N_c$ while in the $SU(3)$ case the
 states belong to the regular representation ${\bf (1,k)}$.
 \par

{}From these results we obtain the $F/D$
ratio of spin-flip baryon vertex in the
$SU(3)$ CSM
\begin{eqnarray}
\left(\frac{F}{D}\right)_{SU(3) \; CSM}
=\frac{N_c^2 + 8 N_c + 27}{3(N_c^2 + 8N_c + 3)} +\cdots.
\end{eqnarray}

In Fig.~1 we compare the $F/D$ ratio in the NRQM with
$SU(2f)$ symmetries and the ratio in the $SU(3)$ CSM.
Two ratios coincide at both $N_c=1$ and
$N_c \to \infty$. The experimental
value of the $F/D$ ratio of spin-flip
baryon vertex $\mu_{\rm exp}
 = 0.65 \pm 0.02$
lies between the two lines nearer to
that of $SU(3)$ chiral soliton.

\section{The $1/N_c$ Expansion of $F/D$ Ratio
in the Nonrelativistic Quark
 model  and Chiral Soliton Model}
\hspace*{\parindent}
If the value of $N_c$ is large enough
we can expand $F/D$ ratios as follows,
\begin{eqnarray}
\left(\frac{F}{D}\right)_{SU(4),SU(6)}
&=& \frac{1}{3} + \frac{4}{3N_c} - \frac{4}{3N_c^2}
+ \cdots \label{fdnrqm} \\
\left(\frac{F}{D}\right)_{SU(3) CSM}
&=& \frac{1}{3} + \frac{0}{N_c}
 + \frac{8}{N_c^2} +
\cdots \label{fdcsm2} \\
\left(\frac{F}{D}\right)_{SU(2) CSM}
&=& \frac{1}{3} +\cdots \label{fdcsm3}
\end{eqnarray}
\hspace*{\parindent}
First we note here  that the limiting
value 1/3 of the $F/D$ ratio is the
consequence of the large $N_c$ counting
rules of the spin-flip baryon vertex.
 Generally the isoscalar part of spin-flip
vertex is $1/N_c$ of the
  isovector part of spin-flip vertex.
Therefore we can neglect the isoscalar
   part in the large $N_c$ limit and we
obtain  the limiting value $F/D=1/3$.
   This is natural from the viewpoint
of the $SU(4)$ and $SU(6)$ symmetric
 NRQMs since $u$ and $d$ quarks are
in the totally symmetric states with
  respect to the spin and isospin and
for the ground state baryons
  the $I=0$ ($I=1$) states
    have $J=0$ ($J=1$) and the spin-flip
transitions are allowed only for
     the isospin-flip transitions.
     Similar situations also occur in the CSM.
     Due to the hedgehog Ansatz $I=J$
structure is realized in the static
     chiral soliton and deviation from
this structure comes from the time
     derivative corrections of dynamical
$SU(2)$ or $SU(3)$ matrix $A(t)$ of
     spin and flavor rotations.
   \par
The second  point to be noted here is
that the $1/N_c$ correction to $F/D$
 ratio has two different origins. One
is the $1/N_c$ correction from baryon
 states and the other is from the $1/N_c$
correction of the dynamical
 quantities. In the $SU(4)$ Model the $1/N_c$
correction comes only from
  baryon states and not from
the vertex operators. Contrary  to the
  $SU(4)$ symmetric model, in the $SU(2)$ CSM
 the $1/N_c$ corrections come
from the dynamical variables
  (time derivative of spin-isospin
rotation matrix $A(t)$) not from the
  baryon states. \par
The third  point is that in the
$SU(4)$ and $SU(6)$ symmetric models
 the $1/{N_c}$
 expansion of $F/D$ ratio (\ref{fdnrqm})
is a convergent expansion for
 $N_c >1$,
 while in the $SU(3)$ CSM the $1/{N_c}$
 expansion (\ref{fdcsm2}) is a divergent
expansion at $N_c= 3$, because the convergence radius
 is $1/N_c =(4-\sqrt{13})/3=1/7.6$ and
$N_c$ must be larger than 8
  for the  convergence
   of expansion.
  Therefore it is not clear whether
the quantitative feature of the chiral
   soliton model with $N_c=3$ is the
same as that  with large enough $N_c$ or
   not.

    \par

{}From the consistency conditions  Dashen,
Jenkins and Manohar
derived $1/N_c$ expansion of ${\cal N}/{\cal M}$
ratio$\cite{Dashen}$,
\begin{equation}
\frac{\cal N}{\cal M} = \frac{1}{2} +
\frac{\alpha}{N_c} + \cdots,
\end{equation}
which  gives the $1/N_c$ expansion of $F/D$ ratio
\begin{equation}
\frac{F}{D} = \frac{1}{3} - \frac{8\alpha}{9N_c} + \cdots.
\end{equation}
In this expansion,  $\alpha$ $= -3/2$
in the $SU(4)$ or $SU(6)$ symmetry
model. On the other hand, in the $SU(2)$
and $SU(3)$ CSM $\alpha$ $= 0$ in
 the static limit.
  \par

\section{Consistency Condition,
$F/D$ Ratios  of $SU(6)$ Symmetry and $SU(3)$
 Chiral Soliton Model}
\hspace*{\parindent}
We study whether the above
results for the NRQMs and the CSMs
  satisfy the consistency conditions
 derived by Dashen, Jenkins
and Manohar$\cite{Dashen}$.
 For simplicity, we consider
here explicitly only the $SU(2)$ part
of the matrix elements of the
axial vector baryon-pion vertex expanding it in
 $1/N_c$ as;
 \begin{equation}
(A^{ia})_{BB'} = N_c (X_0^{ia})_{BB'}
 + (X_1^{ia})_{BB'} + \frac{1}{N_c}
(X_2^{ia})_{BB'} + \cdots .
\end{equation}
The  Gervais-Sakita condition is
\begin{eqnarray}
({\rm I}) \qquad \quad
([X^{ia}_0,X^{jb}_0])_{BB'}=0. \hskip 5.5cm
\end{eqnarray}
Dashen, Jenkins and Manohar
obtain the consistency condition
\begin{eqnarray}
({\rm II}) \qquad ([X^{kc}_0 ,
[X^{jb}_1 , X^{ia}_0]])_{BB'} + ([X^{jb}_0 ,
 [X^{kc}_0 , X^{ia}_1]])_{BB'} = 0. \hskip 1.5cm
\end{eqnarray}
 and
\begin{eqnarray}
  &{\rm (III)}&\quad ([X^{ld}_0,[X^{kc}_0,
[X^{jb}_0,X^{ia}_2]]])_{BB'}
  + ([X^{ld}_0,[X^{kc}_0,[X^{jb}_2,X^{ia}_0]]])_{BB'}
 \hskip 5mm \nonumber \\
& & +([X_0^{ld},[X_0^{kc},[X_1^{jb},X_1^{ia}]]])_{BB'}
    + ([X_0^{ld},[X_1^{kc},[X_1^{jb},X_0^{ia}]]])_{BB'}
 \nonumber \\
& & + [X_0^{ld},[X_1^{kc},[X_0^{jb},X_1^{ia}]]])_{BB'} = 0.
\end{eqnarray}
The consistency condition (III) that we use here
for the sake of simplicity
 is equivalent but has slightly different  expression from
 the one in $\cite{Dashen}$.
\begin{flushleft}
{\bf (i) the NRQMs with the $SU(4)$ and $SU(6)$ symmetries}
\end{flushleft}
The isospin, spin and spin-isospin
operators in the NRQMs are given by
\begin{eqnarray}
& & G^{ia} = q^{\dagger} \sigma^i \tau^a q,  \nonumber \\
& & I^a  = q^{\dagger} \tau^a q,   \nonumber \\
& & J^i = q^{\dagger} \sigma^i q. \nonumber
\end{eqnarray}
where $q^{\dagger}$ and $q$ are
respectively creation and annihilation
 operators.
\par
These operators satisfy $SU(2f)$ algebra
\begin{eqnarray}
& & [I^a,G^{jb}] = i\epsilon_{abc} G^{jc}  \\
& & [J^i,G^{jb}] = i\epsilon_{ijk} G^{kb}  \\
& & [G^{ia},G^{jb}] = \frac{i}{2f}
\epsilon_{ijk} \delta_{ab} J^k +\frac{i}{4}
f_{abc} \delta_{ij} I^c + \frac{i}{2}
\epsilon_{ijk} d_{abc} G^{kc},
 \label{iajb}
 \end{eqnarray}
These operators themselves have no $N_c$
dependence in the $SU(2f)$ symmetric
NRQM. \par
The axial current $A^{ia}$ in QCD can be
expanded by using the NRQM operators
as$\cite{Luty}$
\begin{eqnarray}
  A^{ia} = G^{ia} + \frac{a}{N_c} J^i I^a +\cdots
\end{eqnarray}
{}From large $N_c$ counting rules, $(G^{ia})_{BB'}
\sim O(N_c)$, $(I)_{BB'} \sim
(J)_{BB'} \sim O(N_c^0)$. So the operator $J^i I^a$
 is the higher order term in
 the $1/N_c$ expansion and the breaking term
for the $SU(4)$ ($SU(6)$)
 symmetry.
By taking the matrix element
of $(\ref{iajb})$ between the initial and the
final state baryons $B'$ and  $B$ we obtain
 the following large $N_c$ relations
for the $SU(4)$ symmetric NRQM:
\begin{eqnarray}
& & ([X_0^{ia},X_0^{jb}])_{BB'} = 0 \\
& & ([X_0^{ia},X_1^{jb}])_{BB'} +
([X_1^{ia},X_0^{jb}])_{BB'} = 0 \\
& & ([X_0^{ia},X_2^{jb}])_{BB'} +
([X_2^{ia},X_0^{jb}])_{BB'} +
 ([X_1^{ia},X_1^{jb}])_{BB'} \nonumber \\
& & \;\;\;\;\;\;\;\;\;\;\;\;\; =
\frac{i}{4}\epsilon_{ijk} \delta_{a} (J^k)_{BB'}
 +\frac{i}{4}\epsilon_{abc} \delta_{id} (I^c)_{BB'}
\end{eqnarray}
 Here the contribution from the $d$-term
in the case of $SU(6)$ symmetry
 is absorbed into the isoscalar part.
 \par

{}From the above expressions, it is
obvious that $X_0^{ia}$, $X_1^{ia}$
 and $X_2^{ia}$ satisfy the consistency
conditions (I), (II) and (III).
\begin{flushleft}
{\bf (ii) the $SU(2)$ and $SU(3)$ CSMs}
\end{flushleft}
The isospin, spin and spin-isospin
operators in the CSM are
given by
\begin{eqnarray}
& & {\tilde I}^a =i \lambda {\rm Tr}
(A^{\dagger} \tau^a {\dot A}), \label{ia}
 \\
& & {\tilde J}^i = -i \lambda {\rm Tr}
 (\sigma^i A^{\dagger} {\dot A}),
\label{ji} \\
& & {\tilde G}^{ia}= g_A N_c {\rm Tr}
 (\sigma^i A^{\dagger} \tau^a A),
\end{eqnarray}
respectively where elements of the
 $SU(2)$ or $SU(3)$  matrix $A(t)$
 are
the collective coordinates describing
 spin-isospin rotations, dot means time
derivative. In $(\ref{ia})$ and $(\ref{ji})$
 $\lambda$ is the moment of
 inertia
 of spin-isospin rotation of soliton.
This $\lambda$ is order $N_c$ and
 ${\dot A}$ is order $O(1/N_c)$ thus
 ${\tilde I}^a$ and ${\tilde J}^i$ are
 of order $O(1)$.
These operators satisfy an algebra
\begin{eqnarray}
& & [{\tilde I}^a, {\tilde G}^{jb}] =
i\epsilon_{abc} {\tilde G}^{jc} \\
& & [{\tilde J}^i, {\tilde G}^{jb}] =
i\epsilon_{ijk} {\tilde G}^{kb} \\
& & [{\tilde G}^{ia}, {\tilde G}^{jb}] = 0.
\end{eqnarray}
The axial vertex operator of baryon can be
expanded by using these bases
\begin{equation}
A^{ia} = {\tilde G}^{ia} + \frac{a'}{N_c} {\tilde J}^i
        {\tilde I}^a +\cdots
\end{equation}
{}From large $N_c$ counting rules,
 $({\tilde G}^{ia})_{BB'} \sim O(N_c)$,
 $({\tilde I})_{BB'} \sim ({\tilde J})_{BB'}
 \sim O(N_c^0)$.
  So the operator ${\tilde J}^i {\tilde I}^a$
 is the higher order term in
 the $1/N_c$ expansion and the breaking term
for the contracted $SU(4)$
 ($SU(6)$) symmetry. \par
In the $SU(2)$ CSM case, there are no  $1/N_c$
 corrections
 from baryon states. Thus we obtain
\begin{equation}
(A^{ia})_{BB'} = N_c (X_0^{ia})_{BB'}.
\end{equation}
It is obvious that the $SU(2)$ CSM satisfies
 the consistency
 condition derived by Dashen Jenkins and
 Manohar$\cite{Dashen}$.
  \par
On the other hand, in the $SU(3)$ CSM there are  $1/N_c$
 corrections originating  from baryon states. \par
In this case we obtain the following relations
\begin{eqnarray}
& & ([X_0^{ia},X_0^{jb}])_{BB'} = 0 \\
& & ([X_0^{ia},X_1^{jb}])_{BB'} +
([X_1^{ia},X_0^{jb}])_{BB'}= 0 \\
& & ([X_0^{ia},X_2^{jb}])_{BB'} +
 ([X_2^{ia},X_0^{jb}])_{BB'} +
([X_1^{ia},X_1^{jb}])_{BB'} = 0.
\end{eqnarray}
It is obvious from these expressions
 that $X^{ia}$ satisfy consistency
 conditions (I), (II) and (III) in the $SU(3)$ CSM.
  \par
As is seen from the above analysis the difference
between the NRQM with
 $SU(6)$ symmetry and the $SU(3)$ chiral
soliton model arises from the difference
 of $X_2^{ia}$ in these  two models.
 This is
 consistent with the analysis of Dashen Jenkins
 and Manohar$\cite{Dashen}$.
 The solution of consistency condition  $X_1^{ia}
 \sim X_0^{ia}$ gives
  the limiting value 1/3 for F/D ratio irrespective
 of $X_2^{ia}$ which gives
the  next order correction.
   \par

Here we note that in the $SU(2)$ CSM there are no
 $1/N_c$ corrections from baryon states, but in the
 $SU(3)$ CSM
  baryon states have  $1/N_c$ corrections and give
the $1/N_c$ to
 the matrix elements of the baryon vertices. This is
  the consequence of
 the fact  that the wave functions of proton and
neutron given in the $SU(2)$
  and $SU(3)$ CSMs are different.
  \par
The situation is quite different
in the  NRQM with the $SU(4)$ symmetry and
the $SU(2)$ CSM.
Both in the $SU(4)$ symmetric model and the
 $SU(2)$ CSM the
 nucleons belong to the same fundamental
representation, i.e. the isospin
  doublet and gives the same contributions
  at the leading order in $1/N_c$
   expansion as  discussed by Manohar$\cite{Manohar}$  and
    Bardakci$\cite{Bardakci}$.

\section{Conclusion and Discussions}
\hspace*{\parindent}
We have calculated the $F/D$ ratios
 in the NRQMs with the $SU(4)$ and $SU(6)$
symmetries and the $SU(2)$ and $SU(3)$ CSMs
 to all orders
of $1/N_c$ expansion.
 We have confirmed that the limiting value
 of $F/D$ ratio tends  to 1/3 and
  this result is model independent.

   In general the $1/N_c$ corrections
 to the axial vector coupling of baryons
   have two origins in the model
independent QCD. One is the $1/N_c$
    corrections from baryon states and
the other is the $1/N_c$ corrections
     of operators.   In the $SU(2f)$ symmetric NRQM the
  $N_c$ dependence comes only from the
 wave function of baryon states and not
  from the operators. On the other hand
 in the CSMs the
  operators have $N_c$ dependence and
 the wave functions of the $SU(2)$ model
  have no $1/N_c$ corrections while those
 of the $SU(3)$ model have
   the $1/N_c$ corrections. These model
dependence of $1/N_c$ correction
   are summarized in Table I.
\par
In contrast to the $1/N_c$ expansion
 in the NRQM, we found that the
$1/N_c$ expansion of the $SU(3)$ CSM
 does not converge at
$1/N_c =1/3$. For the $1/N_c$
 expansion of the $SU(3)$ CSM to converge
 the value of $N_c$  must be larger than 8.
  \par
  In connection with convergence problem
  of the $1/N_c$ expansion we note here
  that though the limiting value 1/3 of
$F/D$ ratio can also be derived from
   the analyses of vertices of the other
 baryons  such as $\Sigma$
    and $\Lambda$$\cite{Dashen}$, some of
 the wave functions of large $N_c$
     baryons contain $N_c$ dependence which
 can not be expanded
      at $1/N_c =1/3$.
     For example, if we use transitions
containing the $\Sigma$ and $\Lambda$
  vertices
  \begin{eqnarray}
 \frac{\Sigma^+ \to \Sigma^0 \pi^+}
{\Sigma^+ \to \Lambda \pi^+}
= 1+ O(1/N_c^2) = \sqrt{1+\frac{2}{k}}
 \left(\frac{k {\cal M}-{\cal N}}
 {k {\cal M}+{\cal N}} \right)
\end{eqnarray}
we can derive the correct  limiting
 value 1/3 of the $F/D$ ratio.
But the wave functions of $\Lambda$
 and $\Sigma$ can not be expanded for $N_c
\le 5$.
\par

{\bf Added in proof}:

After this work was completed the second
  paper of Dashen, Jenkins and Manohar
has appeared
$\cite{Jenkins}$  which concerns with  similar problems
 discussed here  but from slightly different  viewpoint.

\eject

\eject
\begin{center}
\begin{tabular}{|c|c|c|} \hline
  Model &\quad Operators \quad &\quad
 Baryon States \quad  \\ \hline\hline
   QCD  &  yes  &  yes   \\ \hline\hline
  \quad $SU(4)$ symmetric NRQM \quad & no  &  yes \\ \hline
  $SU(6)$ symmetric
  NRQM & no  &  yes \\ \hline
  $SU(2)$ CSM  &  yes & no \\ \hline
  $SU(3)$ CSM  &  yes & yes \\ \hline
  \end{tabular}
\vskip 5mm
Table I. $N_c$ dependence of operators and baryon states
\end{center}

\vskip 3cm

{\large \bf Figure Caption}

\begin{flushleft}
\vskip 1cm
Fig. 1.\quad $F/D$ ratios in
 the $SU(6)$ symmetric NRQM(lower curve) and
$SU(3)$ CSM(upper curve). The large $N_c$ limiting
 value 1/3 is shown by
an arrow and the experimental
 value by a bullet at $N_c=3$.
\end{flushleft}

\end{document}